\documentclass[aps,prd,preprint,nofootinbib]{revtex4}

\usepackage{graphicx}
\usepackage{dcolumn}
\usepackage{bm}
\usepackage{amssymb}
\usepackage{amsmath}
\usepackage{epsfig}
\newcommand{\newc}{\newcommand}
\newc{\beq}{\begin{equation}}
\newc{\eeq}{\end{equation}}
\newc{\eeqa}{\end{eqnarray}}
\newc{\beqa}{\begin{eqnarray}}
\newc{\bsym}{\boldsymbol}
\newc{\mrm}{\mathrm}
\newc{\ovl}{\overline}
\newc{\ovla}{\overleftarrow}
\newc{\ovra}{\overrightarrow}
\newc{\ra}{\rightarrow}
\newc{\lra}{\leftrightarrow}
\newc{\wtil}{\widetilde}
\newc{\eps}{\epsilon}
\newc{\hc}{\dagger}
\newc{\pd}{\partial}
\newc{\SL}{\!\!\!/}
\newc{\LH}{\hat{L}}
\newc{\RH}{\hat{R}}
\newc{\sWsq}{\sin^2\theta_\mathrm{W}}
\newc{\cWsq}{\cos^2\theta_\mathrm{W}}
\newc{\half}{\frac{1}{2}}
\newc{\hh}{\hat{H}}
\newc{\hphi}{\hat{\Phi}}
\newc{\nonr}{\nonumber}

\begin{document}
\title{A model for Neutrino Masses and Dark Matter with the Discrete Gauge Symmetry}

\author{We-Fu Chang}
\email{wfchang@phys.nthu.edu.tw}
\author{ Chi-Fong Wong}

\affiliation{Department of Physics, National Tsing Hua University,
HsinChu 300, Taiwan}

\begin{abstract}
A simple renormalizable $U(1)$ gauge model is constructed to
explain  the smallness of the active neutrino masses  and provide
the stable cold dark matter candidate simultaneously. The local
$U(1)$ symmetry is assumed to be spontaneously broken by a scalar
field around the TeV scale. The active neutrino masses are then
generated at one-loop level. This model contains several cold dark
matter candidates whose stability is guaranteed  by a  residual
discrete gauge $Z_2$ symmetry a la the Krauss-Wilczek mechanism.
Unlike the other dark matter models, no further global discrete or
continuous symmetry is introduced. Moreover, the masses of all  fermionic
degrees of freedom beyond the Standard Model are closely related to
the scale of spontaneous breaking of $U(1)$ thus they could be probed at or below the TeV scale.
The possible cosmological and  phenomenological consequences are briefly discussed.

\end{abstract}
\maketitle
\section{Introduction}
It is now widely accepted that at least two kinds of neutrinos are
massive\cite{PDG}. The pressing questions we are facing now: (1)  how many extra degrees
of freedom beyond the Standard Model (SM) are responsible for
generating the neutrino masses, and (2) are they fermionic or bosonic?
On the phenomenological side, it is practical to ask whether we
are able to test the neutrino mass generation mechanism directly.
If the new degrees of freedom decouple at a rather high energy
scale, unfortunately this happens for some of the cases,  we are
left with only one effective dim-5 operator $(L \Phi)^2$ at  low
energy. For example, there is no way to directly test the usual
seesaw mechanism\cite{seesaw} with $\sim 10^{14}$ GeV or higher Majorana
neutrinos. And we can only study its indirect consequences, for
example, the leptogenesis.
It will be interesting, although unnecessary, to have a neutrino mass generation mechanism
 and at the same time
the responsible new degrees of freedom are accessible within
human's reach.  It was first pointed out by Zee and
Babu\cite{radi_nuM} that the active neutrino masses could be
generated radiatively. In that case, due to the loop suppression
factors, the active neutrino masses are naturally small compared
to other SM charged fermions. Although the masses of the new
degrees of freedom are  totally arbitrary, they are more likely to
be probed experimentally in some parameter space. Therefore, we
aim for a model in which the neutrino masses are radiatively
generated\footnote{ For the follow ups and other proposals for
generating neutrino mass, readers may want to consult a recent
review\cite{Chen:2011de}.  }. Moreover, we hope to have a unified
mechanism such that  the mass scale of the new degrees of freedom
is not totally arbitrary.

 On the other
hand,  there are several ways to infer the dark matter abundance
$\Omega_{DM}$ in the Universe\cite{DarkMatter}. The existence of
dark matter with $\Omega_{DM} h^2\sim 0.11$, where $h$ is the
Hubble constant in units of 100 km/(s.Mpc), requires physics
beyond the SM. It will be even more interesting if  one of
the new fields for radiatively generating neutrino mass serves as
the dark matter candidate. In fact, several interesting models
have been proposed to connect the origin of neutrino masses to the
existence of the cold dark matter\cite{NuDMold,Kubo:2006rm}.
And, a viable dark matter candidate  must be as long-lived as the
Universe. It is common for people to impose global discrete or
continuous symmetry to stabilize the dark matter candidate from
decaying ( \cite{Kubo:2006rm} is an exception, but it failed to
 come up with a  theory which is renormalizable and free of anomalies. )
  However, the origin of these global discrete or continuous symmetries is not explained.
Moreover,  quantum gravity effects do not respect global symmetries~\cite{Krauss:1988zc}.
One elegant remedy  is to use the  discrete symmetry originated from a spontaneously broken gauged symmetry,
known as the Krauss-Wilczek mechanism(KWM)\cite{Krauss:1988zc}. For a recent implementation of KWM
to stabilize the dark matter candidate see \cite{Batell:2010bp} and references therein.

In this paper, a $U(1)_\nu$ gauge symmetry is introduced on top of
the SM interactions and it is responsible for all three features
mentioned above: neutrino masses generation, testable new degrees
of freedom, and the existence of stable dark matter. The
$U(1)_\nu$  is spontaneously broken by a SM singlet scalar which
carries two units of $U(1)_\nu$ charge. A gauged discrete $Z_{\nu
2}$ symmetry can remain after the spontaneous symmetry breaking
(SSB) of $U(1)_\nu$ and the $Z_{\nu 2}$-odd cold dark matter
candidate can be stable.  While each ingredient is not new, to
our best knowledge, ours is the first successful model which
conjoins all.

\section{Model Setup}
\label{sec:model}
 We present a minimal model which makes use of
the KWM and radiatively generates the neutrino masses at the
lowest possible mass dimension. In addition to the SM fermions and
Higgs doublet $\Phi$, this model consists of two extra pairs of
chiral fermions $N_{R_{1,2}}$ and $n_{L_{1,2}}$, one extra Higgs
doublet  $\eta$, and two complex scalars $S$ and $\sigma$. Their
quantum numbers are summarized in Table \ref{tab_charge}. The
$U(1)_\nu$ charge assignment forbids the $N_{Ra}$ and $n_{La}$ to
have Majorana masses. Note that we need at least two pairs of
$N_R$ and $n_L$ to accommodate the observed neutrino data.
Introducing vector-like fermions is needed for cancelling the
anomalies.
\begin{table}[ht]
  \centering
  \begin{tabular}{|c|ccc|cc|cc|cccc|}
\hline
  & $Q_{Li}$& $u_{Ri}$ & $d_{Ri}$& $L_i$ & $e_{Ri}$ & $N_{Ra}$ & $n_{Lb}$ & $\Phi$ & $\eta$ & $\sigma$ & $S$   \\
\hline
$SU(2)_L$ &$2$ & $1$& $1$& $2$ & $1$& $1$& $1$& $2$& $2$& $1$& $1$\\
$U(1)_Y$& $\frac 16$ & $ \frac 23$ & $-\frac 13$& $-\frac 12$ & $-1$ &$0$ &$0$ &$\frac12$ & $\frac12$&$0$&$0$\\
\hline
$U(1)_\nu$& $0$ &$0$ &$0$ & $0$ & $0$ & $-1$ & $-1$ & $0$&$-1$& $-1$ & $2$\\
\hline
$Z_{2\nu}$& $+$ &$+$ &$+$ & $+$ & $+$ & $-$ & $-$ & $+$&$ -$&$-$ & (N.A.) \\
\hline
\end{tabular}
  \caption{Charge assignment and the remaining  discrete $Z_{2\nu}$ parity for the fields,
where $i=1,2,3$, $a,b=1,2$, and $Q_L, u_R,d_R, L, e_R$ are the standard notation for the SM quark and lepton.
  }\label{tab_charge}
\end{table}
In this model, the $U(1)_\nu$ invariant Yukawa couplings and the Dirac mass term
between $N_{Ra}$ and $n_{Lb}$ are:
\beq
 \frac{y^N_a}{2} \overline{N^C_a} S N_a
+ \frac{y^n_a}{2} \overline{n^C_a} S n_a +g_{ia}
\overline{L_i}\tilde{\eta} N_a + m^D_{ab}\bar{n}_a N_b + h.c. \eeq
If the first two terms were absent ( or $y^{N,n}=0$), a global
axial $U(1)_A$ symmetry which transforms $N_R\ra e^{i\theta}N_R$
and $n_L\ra e^{-i\theta}n_L$ will forbid the Dirac mass term. The
presence of the first two terms explicitly violates the $U(1)_A$
symmetry and thus suggests a natural scale for $m^D_{ab}
\sim y \langle S\rangle$, although it is perfectly legitimate for
$m^D_{ab}$ to take ANY other value. In this work, we simply
focus on the scenario that $m^D$ is around TeV from the standpoint
of being phenomenologically interesting.  The Yukawa couplings
$y^N$ and $y^n$ can be taken to be diagonal without losing any
generality. The most general renormalizable scalar potential in
this model is
\beqa
V&=& \bar{\mu}_\Phi^{2} |\Phi|^2 + \bar{\mu}_\eta^2 |\eta|^2 +\bar{\mu}_\sigma^2 |\sigma|^2 +\bar{\mu}_S^2 |S|^2 \nonr\\
&+& \bar{\lambda}_1 |\Phi|^4 +\bar{\lambda}_2 |\eta|^4+\bar{\lambda}_3 |\sigma|^4+\lambda_4 |S|^4\nonr\\
&+&\bar{\lambda}_5 |\Phi|^2|\eta|^2  +\lambda_6 |\Phi^\dag \eta|^2+\bar{\lambda}_7 |\Phi|^2|\sigma|^2+\lambda_8 |\Phi|^2|S|^2\nonr\\
&+&\bar{\lambda}_9 |\eta|^2|\sigma|^2 +\lambda_{10} |\eta|^2|S|^2+\lambda_{11} |\sigma|^2|S|^2\nonr\\
&+& \kappa(\Phi^\dag \eta \sigma S) + \mu_1 (\sigma\sigma S ) +\mu_2 (\eta^\dag \Phi \sigma) + h.c.
\label{Eq:fullV}
\eeqa
The parameters $\kappa, \mu_{1,2}$ can be taken to be real positive.
However, this potential is too complicated for one to obtain any meaningful constraint on the parameters.
Instead of going for a full analysis of Eq.(\ref{Eq:fullV}), we assume that $S$ gets a positive real vacuum expectation value (VEV),
 $v_S \sim TeV$. We assume that the mass of $S$, $( =\sqrt{ -2 \bar{\mu}_S^2 })$,  is around TeV as well, such that $\lambda_4 = -\bar{\mu}_S^2 /2 v_S^2 \leq 1$
 can be met and the scalar sector is still  perturbative.
When the energy scale is less than $\Lambda= \sqrt{ -2
\bar{\mu}_S^2 }\sim$ TeV, the degrees of freedom of $S$ are
integrated out and we work with an effective theory without $S$.

 The covariant derivative of $S$ is given by
$ D_\mu S = \left(\partial_\mu - i g_\nu X_\mu \right) S $,
where $X_\mu$ is the $U(1)_\nu$ gauge field, and $g_\nu$ the gauge coupling constant which
should take a value similar to the SM ones.
 After SSB, $S$ is parametrized as
$S = (v_S + S_R) \exp(i\tau_S/v_S)$. The Goldstone field,
$\tau_S$, can be removed by a gauge transformation $X_\mu \ra
X_\mu - (\partial_\mu\tau_S / g_\nu v_S)$ accompanied by a
concomitant redefinition of all other fields, $f$, which carry
$U(1)_\nu$ charge $Q_f$: $f \ra f\exp\left(-i \tau_S Q_f/ 2 v_S
\right)$. It is clear that a transformation $\tau_S \ra \tau_S +
2\pi v_S$ leaves the background vacuum configuration invariant.
The $Z_{\nu 2}$ symmetry then emerges  since  the charges of
all  fields other than $S$   are either $\pm 1$ ($Z_{\nu 2}$-odd)
or $0$ ($Z_{\nu 2}$-even) under the $U(1)_\nu$.

 The gauge boson of $U(1)_\nu$, dubbed $Z_\nu'$,  thus gets a mass $\sqrt{2} g_\nu v_S \lesssim $  TeV.
The singlet fermions $N_a$ and $n_a$ now acquire their Majorana masses $( y^N_a v_S)$ and $(y^n_a v_S)$ respectively.
Four Majorana states $\chi_{1-4}$ can be constructed from $N_a$ and $n_a$ and their charge conjugate
by diagonalizing a $4\times 4$ mass matrix. Since we do not attempt to fit the neutrino oscillation data in this letter,
the explicit form of the mixing is not our concern here. The bottom
line is that the mixings between the $n_L$ and  $N_R$ sectors are order one, $\tan 2\theta_{Nn}\sim m^D/v_S(y^N  -y^n)$.
We denote the lightest(heaviest)  mass eigenstate as $\chi_1(\chi_4)$.

\begin{figure}[htb]
\centering
\includegraphics[width=5in]{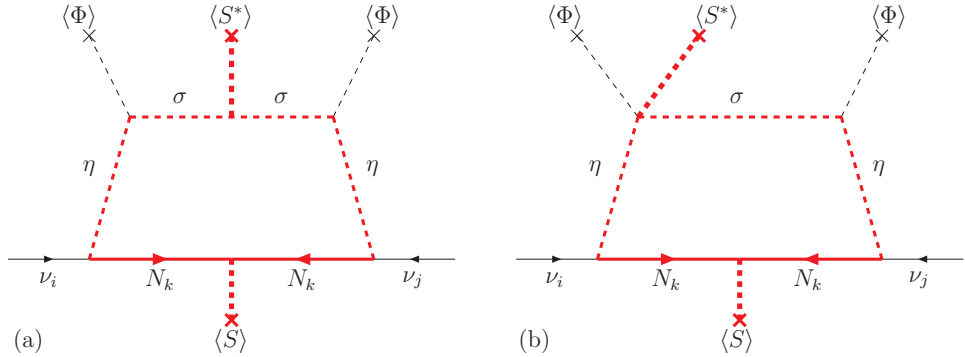}
\caption{ The 1-loop diagrams which give rise to the active
neutrino masses.  The mirror image of  diagram-(b) which has the $\langle S^*\rangle$ attached to
the $\sigma\eta\langle\Phi\rangle$ vertex on the right-hand side is not shown.
Those charged under $U(1)_\nu$ are the red(thick) lines.
\label{fig:1loop} }
\end{figure}

The active neutrinos receive their Majorana masses via the one-loop diagrams displayed in Fig.\ref{fig:1loop}.
From the effective operator point of view, the active neutrino masses are attributed to  a dimension-seven operator
$(\Phi L)^2(S^\dag S)$.
The resulting neutrino mass matrix element $(m_\nu)_{ij}$ from Fig.-\ref{fig:1loop}(a) and Fig.-\ref{fig:1loop}(b) can be estimated to be
\beq
 (m_\nu)^{(a)}_{ij}\sim { \mu_1 \mu_2^2 v_\Phi^2 v_S^2 \over 16\pi^2\Lambda^6}
\sum_a  y_a^N g_{ia}^*g_{ja}^* \,,\; \mbox{and}\;\;
 (m_\nu)^{(b)}_{ij}\sim {\kappa \mu_2  v_\Phi^2 v_S^2 \over 16\pi^2 \Lambda^4}
\sum_a  y_a^N g_{ia}^*g_{ja}^*
\eeq
respectively, where
$\langle \Phi \rangle=v_\Phi= 174$GeV is the VEV of the SM Higgs.
If taking the dimensional couplings $\mu_1$ and $\mu_2$ to be electroweak $\sim 0.1 \Lambda$,
then $ m_\nu^{(b)}/ m_\nu^{(a)} \sim  (\kappa  \Lambda^2 / \mu_1\mu_2)\gg 1$ and Fig.-\ref{fig:1loop}(b) gives the dominant contribution
to active neutrino masses
\beq
\sim 0.01\times
\frac{|g|^2}{16\pi^2}  \kappa  y^N \mu_2  \,.
\label{eq:NuM1lp}
\eeq
In Eq.(\ref{eq:NuM1lp}), if we take a rather
conservative estimation  that $ \kappa y^N \sim 0.1 $,
 $\mu_2 \sim 100$ GeV,  and Yukawa coupling
$|g|\sim 10^{-4}$ ( roughly 10 times of the Yukawa
coupling for the SM electron) the resulting neutrino masses are in the sub-eV range without much fine-tuning.
Moreover, if  $m^D\sim y v_S$, the
Majorana states, $\chi$'s, are still around but slightly below the TeV range
and could be studied by the man-made machines.

Note that there are two-loop contributions to active neutrino masses if  the two external $S$-legs
in Fig.-\ref{fig:1loop} are connected. And those two-loop contributions  are equivalent to the dimension-five
operator $(\Phi L)^2$ in low-energy effective theory. With the help of dimensional analysis, their magnitudes can be estimated to be
\beq
 (m_\nu)^{(a)}_{ij,\mbox{2-loop}}\sim { \mu_1 \mu_2^2 v_\Phi^2  \over (16\pi^2)^2\Lambda^4}
\sum_a  y_a^N g_{ia}^*g_{ja}^* \,,\; \mbox{and}\;\;
 (m_\nu)^{(b)}_{ij,\mbox{2-loop}}\sim {\kappa \mu_2  v_\Phi^2  \over (16\pi^2)^2 \Lambda^2}
\sum_a  y_a^N g_{ia}^*g_{ja}^*
\eeq
respectively. Again, the connected diagram-\ref{fig:1loop}(b) gives dominant two-loop contribution to neutrino masses.
In a perturbative theory, the one-loop diagrams are usually much more important than the two-loop diagrams.
However, in this model, the one-loop contributions vanish before the SSB of $U(1)_\nu$. At that $U(1)_\nu$-symmetric
stage, the active neutrino masses are controlled by the two-loop diagrams where the cutoff should be replaced by
the masses of heavy scalars running in the loop. In some interesting scenarios where $(v_S/\Lambda)^2 \leq 1/(16\pi^2) $, the active
neutrino masses are also governed by the two-loop contributions. In those cases,  the  gauge boson $Z_\nu'$,
the neutral scalar $S_R$ (if the theory is still perturbative ), and some of the Majorana states, $\chi$'s, are all light, $\sim v_S$, and make the terrestrial experimental
probes more probable. But the following effective theory treatment will not be suitable for a light $S_R$ and we
leave detailed exploration of this direction for future study.

 To simplify the discussion,  below $\Lambda$, we assume that
$S_R$ decouples from the rest and yields an effective potential:
\beqa V_{eff} \simeq   \mu_\Phi^2 |\Phi|^2 +\mu_\eta^2 |\eta|^2 +
\mu_\sigma^2 |\sigma|^2
+\lambda_1 |\Phi|^4 +\lambda_2 |\eta|^4 \nonr\\
+\lambda_3 |\sigma|^4+\lambda_5 |\Phi|^2|\eta|^2  +\lambda_6 |\Phi^\dag \eta|^2+\lambda_7 |\Phi|^2|\sigma|^2  \nonr\\
+\lambda_9 |\eta|^2|\sigma|^2+ \kappa v_S (\Phi^\dag \eta \sigma )
+ \mu_1 v_S (\sigma\sigma ) +\mu_2 (\eta^\dag \Phi \sigma) + h.c.
\label{eq:Veff} \eeqa where $ \mu_\Phi^2 = (
\bar{\mu}_\Phi^2+\lambda_8 v_S^2)$, $\mu_\eta^2 = (
\bar{\mu}_\eta^2+\lambda_{10} v_S^2)$ and $\mu_\sigma^2 = (
\bar{\mu}_\sigma^2+\lambda_{11} v_S^2)$.
The  quartic couplings also receive  contributions from tree-level $S$ exchange diagrams:
\beqa
\lambda_1 &=& \bar{\lambda}_1 +{\cal O}(\lambda^2_8)\,,\nonr\\
\lambda_2 &=& \bar{\lambda}_2 +{\cal O}(\lambda^2_{10})\,,\nonr\\
\lambda_3 &=& \bar{\lambda}_3 +{\cal O}(\lambda^2_{11})+ {\cal O}\left(\frac{\lambda_{11} \mu_1}{\Lambda}\right)+ {\cal O}\left(\frac{\mu_1^2}{\Lambda^2}\right)\,,\nonr\\
\lambda_5 &=& \bar{\lambda}_5 +{\cal O}(\lambda_8\lambda_{10})\,,\nonr\\
\lambda_7 &=& \bar{\lambda}_7 +{\cal O}(\lambda_8\lambda_{11})+ {\cal O}\left(\frac{\lambda_8 \mu_1}{\Lambda}\right)\,,\nonr\\
\lambda_9 &=& \bar{\lambda}_9 +{\cal O}(\lambda_{10}\lambda_{11})+ {\cal O}\left(\frac{\lambda_{10} \mu_1}{\Lambda}\right)\,.
\eeqa
Since we assume that $\mu_1\sim 0.1 \Lambda$ and our model is in the perturbative region as well, these effects and the explicit values of $\lambda$'s
are not important to our current
discussion.
This potential is bounded
from below if all $\lambda$'s are set to be positive. The $Z_{\nu
2}$ symmetry will be broken if any of $\eta$ or $\sigma$ develop
nonzero VEV, but it is straightforward to verify that the true
minimum solution that $\langle \Phi \rangle
=v_\phi=\sqrt{-\mu_\Phi^2/2\lambda_1}$ and $\langle \eta \rangle
=\langle \sigma \rangle=0$ can be easily accommodated in
Eq.(\ref{eq:Veff}). For simplicity, here we do not consider
spontaneous  CP violation and the nonrenormalizable operators
which are irrelevant to the current study.

Both the mixing between $\Phi$ and  $\eta$ or $\sigma$ and
the mixing between the SM neutrinos and  $\chi$'s  are forbidden by the $Z_{2\nu}$ symmetry.
There is only one $Z_{2\nu}$-even physical scalar field which is identified as the the SM Higgs $h^0$,
and  six ( 2 charged, 2 scalar, and 2 pseudoscalar ) extra $Z_{2\nu}$-odd physical Higgs.
This feature is very different from the other multi-Higgs models. The  masses of the $Z_{2\nu}$-odd
 Higgs are free but expected to be around sub-electroweak to TeV scale. Due to the $Z_{2\nu}$ parity, the
 sterile fermions in this model do NOT mix with the active neutrinos; therefore  this model cannot explain
 the LSND/MiniBoone anomalies\cite{Nelson:2010hz}.

\section{Phenomenology}
  For the $Z_{2\nu}$-odd scalar sector, it is
easy to work out the  mass square  $M_\pm^2 = \mu_\eta^2+\lambda_5 v_\Phi^2$ for the charged Higgs,
and the mass square matrices  for the scalar and pseudoscalar bosons are
\beqa
M^s_{odd} &=& \left( \begin{array}{cc}
M_\pm^2 +\lambda_6 v_\Phi^2 & \mu_2 v_\Phi +\kappa v_S v_\Phi \\
\mu_2 v_\Phi +\kappa v_S v_\Phi & \mu_\sigma^2+\lambda_7 v_\Phi^2 +2\mu_1 v_S
\end{array}\right)\,,\nonr\\
M^p_{odd} &=& \left( \begin{array}{cc}
M_\pm^2 +\lambda_6 v_\Phi^2 & \mu_2 v_\Phi  - \kappa v_S v_\Phi\\
\mu_2 v_\Phi - \kappa v_S v_\Phi &  \mu_\sigma^2+\lambda_7 v_\Phi^2 - 2\mu_1 v_S
\end{array}\right)\,,
\eeqa
in the basis of $\{\mbox{Re}\,\eta^0, \mbox{Re}\,\sigma^0 \}$ and  $\{\mbox{Im}\,\eta^0, \mbox{Im}\,\sigma^0 \}$ respectively.
For the convenience of latter discussion, we denote the mass eigenstates of physical scalar/pseudoscalar
as $H_{1,2}$/$A_{1,2}$,  and the subscript 1/2 stands for the lighter/heavier ones.  We parametrize the mixing
 angles $\alpha$ and $\delta$ as $H_1 =\cos \alpha \,\mbox{Re}\,\eta^0 +\sin\alpha \,\mbox{Re}\, \sigma^0$
 and $A_1 =\cos \delta\, \mbox{Im}\,\eta^0 +\sin\delta\, \mbox{Im}\,\sigma^0$.
Their masses should be naturally around $v_\Phi$ to $v_S$. However, the fine-tuned  case
that light $H_1$ or $A_1$ is around few GeV cannot be ruled out.
Which $Z_{\nu 2}$-odd degree of freedom
is the  viable dark matter candidate will be made clear soon.

Nonzero neutrino mass  implies that the lepton flavor is no longer
conserved. The null result in searching for one such process sets
a stringent upper limit $Br(\mu\ra e \gamma)<1.2\times 10^{-11}$
\cite{PDG}. The  contribution  from  active neutrinos is highly
suppressed by the Glashow-Ilipoulos-Maiani  mechanism so the
process is dominated by the new physics. The $\mu\ra e \gamma$
like transition is attributed to a loop-generated dimension-six
operator $\bar{L} \Phi \sigma^{\mu\nu} e_R F_{\mu\nu}$. And the
$\mu\ra e \gamma$ branching ratio ( normalized to $\mu\ra
e\bar{\nu}_e\nu_\mu$) can be estimated to be \beq \sim \left( e |
g_{\mu k}^* g_{ek}| v_\Phi\over (16\pi^2) G_F \Lambda^3 \right)^2
\sim 10^{-8}\times |g|^4 \times \left( { 1 \mbox{TeV} \over
\Lambda}\right)^6 \,, \eeq where $g^2$ represent a general Yukawa
coupling product in the one-loop diagram. The $\mu\ra e \gamma$
process  does not impose further constraint because in our
scenario   $|g| \sim 10^{-4}$ is set by the neutrino masses. The
new physics impact on the neutrinoless double beta decay and
$a_\mu$ is insignificant due to the residual  $Z_{\nu 2}$ parity.

The gauge boson $Z'_\nu$  can couple to the SM sector through a
kinetic mixing  term $-\frac{\epsilon}{2} B^{\mu\nu}X_{\mu\nu}$,
where $B^{\mu\nu}(X^{\mu\nu})$ is the SM hypercharge ($U(1)_\nu$) field strength.
The analysis of \cite{Chang:2006fp} works perfectly for this
model. By taking $\epsilon \sim 0.07$, which makes the global electroweak precision tests fit
worsen by $1\%$, it was shown that the Drell-Yan production of a TeV range $Z'_\nu$ at LHC is possible.
Moreover, the TeV range $Z'_\nu$ has  definite relative decay branching ratios into the SM fermions which are
determined completely by the hypercharge of fermion and the mixing parameter $\epsilon$\cite{Chang:2006fp}:
$B(Z'_\nu\ra u\bar{u}): B(Z'_\nu\ra d\bar{d}): B(Z'_\nu\ra e\bar{e}) : B(Z'_\nu\ra \nu\bar{\nu})
= 5.63:  1.66: 4.99:1$.
However, in the case that  $\chi_1$ and $H_1, A_1$ are much lighter than $Z'_\nu$, $Z'_\nu\ra \chi_1 \chi_1, H_1 H_1, A_1A_1$ will
become the dominant decay channels.

\section{Cosmological consequences}
\begin{figure}[htb]
\centering
\includegraphics[width=3.5in]{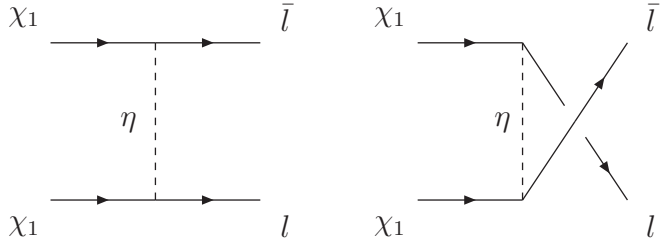}
\caption{ The leading contribution(shown in the interaction basis)
to the annihilation of the lightest $Z_{\nu2}$-odd Majorana
fermion $\chi_1$. \label{fig:N_DM_ann} }
\end{figure}

In this model, $\chi_1, H_1$, and $A_1$ are the  potential cold
dark matter candidates. If $\chi_1$ is the lightest $Z_{2\nu}$-odd
state, it annihilates into the SM leptons through the tree-level t- and
u-channel diagrams mediated by $\eta$, see Fig.\ref{fig:N_DM_ann}.
The annihilation cross section is given by (ignoring the SM lepton
masses )
\beq
\sigma_{ann} v_{rel} = \frac{v_{rel}^2 }{24\pi
M_\chi^2} \sum_{i j}|g_{i1}g^*_{j 1}|^2 x^2(1-2x+2x^2)\,,
\eeq
where $x= M_\chi^2 /(M_\eta^2 +M_\chi^2)$ and  $i,j$ stand for the
final state lepton. However, given that the  Yukawa coupling
$\sim10^{-4}$,
 $M_\eta \ll M_\chi$ is required \cite{N_DM} to yield the relic density of $\Omega_{\chi_1} h^2\sim 0.11$ \cite{DarkMatter},
which contradicts the assumption that $\chi_1$ is the dark matter candidate.
Since that $M_{\chi_{1-4}}> M_\eta$, now we have to check that the
densities of the four Majorana states diminish quickly as the
universe cools down.

\begin{figure}[htb]
\centering
\includegraphics[width=2in]{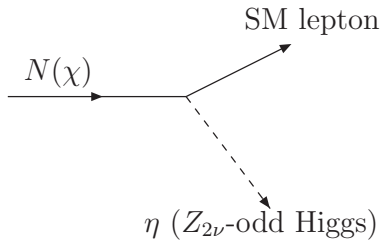}
\caption{ The Feynman diagram in the interaction basis for $N$ to decay.
The mass states are indicated in the parentheses.
\label{fig:NDecay} }
\end{figure}

In the interaction basis, the decays of $\chi$'s happen via the Yukawa coupling
$ g_{ia}\bar{L}_i \tilde{\eta}N_a +h.c.$  and the mixing between
the $n_L$ and $N_R$ sectors, see Fig.\ref{fig:NDecay}.
The final state SM lepton can be treated massless and the decay width can be carried out
straightforwardly:
\beq \Gamma_\chi =\frac{|g|^2
}{16\pi}\times \sin^2\theta \times M_{\chi} \left( 1- \frac{m_\eta^2}{M_\chi^2}
\right)^2
\eeq
where  $\sin\theta$ represents all the mixing
between the mass eigenstates and interaction eigenstates, also the
inexplicit summation over all possible kinematically allowed
channels is understood.   Due to the working assumption that
$M_D\sim y v_S$ the mixing between $\chi_{1-4}$ and $N$ are about
order one, i.e. $\sin^2\theta \sim {\cal O}(1)$, as discussed in
Sec.\ref{sec:model}. The decay rate is to be compared with  the Hubble
constant when temperature is around $\sim M_\chi$
\beq
H(M_\chi) =
\sqrt{{4\pi^3 g_* \over 45}} \frac{M_\chi^2}{M_{Planck}}\,,
\eeq
where $g_*\sim 100$ is the effective number of degrees of freedom
at temperature $\sim M_\chi$. With $|g|^2 \sim 10^{-8}$, $m_\eta /
M_\chi \sim {\cal O}(1)$, we find
\beq
\label{eq:equiv} {
\Gamma_\chi \over H(M_\chi) } \sim \left( 3.7\times 10^5 \right)
\cdot \sin^2\theta \cdot \left( 1- \frac{m_\eta^2}{M_\chi^2}
\right)^2 \cdot \left( { 1 \mbox{TeV} \over {M_\chi}} \right) \gg
1\,.
\eeq
We conclude that the Majorana states decay into the SM final
states and  parity odd scalars fast enough as the Universe
expands.

\begin{figure}[htb]
\centering
\includegraphics[width=4in]{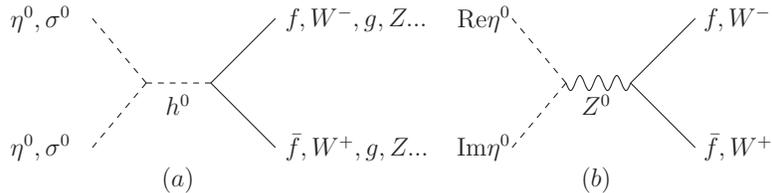}
\caption{ The leading contribution (shown in the interaction basis) to the annihilation and coannihilation of the lightest $Z_{\nu2}$-odd scalar.
\label{fig:S_DM_ann} }
\end{figure}

Therefore,  either $H_1$ or $A_1$ is the viable dark matter candidate and all the heavier $Z_{2\nu}$-odd
scalars decay into the SM $W^\pm/Z^0$ plus $H_1$ or $A_1$.
We use $M_S$ to denote the lighter one of  $M_{H_1}$ and $M_{A_1}$. For $M_S< m_{h^0}$, the leading order
contribution to the scalar dark matter annihilation is through the SM Higgs exchange, see Fig.\ref{fig:S_DM_ann}(a).
The annihilation cross section is given by \cite{Burgess:2000yq}
\beq
\sigma_{ann} v_{rel} ={ 8 \lambda^2 v_\Phi^2 \sum_i \Gamma(h^0\ra X_i)\over (4 M_S^2-m_{h^0}^2)^2 +\Gamma^2_{h^0} m_{h^0}^2 }
{ 1\over 2 M_S}\,,
\eeq
where $\lambda = \cos^2\alpha(\lambda_5+\lambda_6)+\sin^2\alpha \lambda_7
+ \sin 2\alpha(\mu_2+\kappa v_S)/v_\Phi $
 for $H_1$,
and $\lambda = \cos^2\delta(\lambda_5+\lambda_6)+\sin^2\delta\lambda_7+\sin 2\delta (\mu_2 -\kappa v_S)/v_\Phi$ for $A_1$.
Here $\Gamma_{h^0}$ is the total SM Higgs decay width, and $\Gamma(h^0\ra X_i)$ is the
partial rate for the virtual Higgs with mass around $2M_S$ which decays into  $X_i$.
In the limit that $M_S\gg m_{h^0}$, the dominant final states are $W^+ W^-, ZZ$ and $h^0 h^0$ and the annihilation cross section
becomes $\sigma_{ann}v_{rel} \sim \lambda^2/(4\pi M_S^2)$. When the masses  of $H_1$ and $A_1$ are not too different from each other,
one needs to include the coannihilation process, see Fig.\ref{fig:S_DM_ann}(b).
A great deal of effort has been devoted to different aspects of the scalar dark matter including the relic density,
its production and detection at the colliders, and its direct detection at various underground laboratories, see
\cite{S_DM,Burgess:2000yq,SDM_coll}.
In short, all the studies agree that
the scalar dark matter is viable and could be directly detected at the underground laboratories in the near future.
However, to have the right dark matter relic density,  $M_S$ and  $\lambda$
are strongly correlated and such tight relation does not naturally come out in the general scalar
dark matter models, and neither does it in this model.

The usual leptogenesis mechanism does not work in this model. The Yukawa coupling  in our model
 is too large  such that out of equilibrium condition
 cannot be met, see Eq.(\ref{eq:equiv}).
To utilize the TeV scale singlet fermions  for leptogenesis
requires extra arrangement such as the resonance
leptogenesis\cite{Pilaftsis:2003gt} or via the three body decay
mechanism\cite{Hambye:2001eu}. But, some degree of fine-tuning is
then unavoidable. We note in passing that the $Z_{2\nu}$-odd
scalar sector still helps to get a stronger first order
electroweak phase transition which is crucial for successful
electroweak baryogenesis.

\section{Conclusion and discussion}
We construct a simple $U(1)_\nu$ gauge model to address the active
neutrino mass generation, the testable new degrees of freedom
beyond the SM, and the cold dark matter candidate at the same time.
The active neutrino masses arise from one-loop diagrams, the effect is equivalent to
a dim-7 operator at  low energy, without much fine-tuning. The cold dark matter candidate is protected from decaying
by a $Z_{\nu 2}$ parity a la KWM without extra global
discrete or continuous symmetry introduced. The thermal relic density of the
lightest $Z_{\nu 2}$-odd scalar can explain the observed dark
matter abundance, albeit  fine-tuning is required. All new
fermions' masses are related to the SSB of $U(1)_\nu$ thus they
can be probed at or below  TeV scale. The lightest scalar and
pseudoscalar  can be pair produced associated with the SM Higgs
through the $h^0 H_1 H_1, h^0 A_1 A_1$ vertices, or $H_1 A_1$ can
be produced via the $Z^0 H_1 A_1$ coupling. For the charged Higgs,
it can be produced at the LHC via $pp\ra W^{\pm*}\ra H_1 H^\pm,
A_1 H^\pm$ or $pp\ra \gamma^*/Z^{0*}\ra H^\mp H^\pm$. The lightest
Majorana fermion $\chi_1$ is most likely to be studied via the
$U(1)_\nu$  gauge boson decay. Finally, the mass and mixing
pattern in the lepton sector is not explained in this model. It
could be purely accidental, due to some flavor
symmetry\cite{Ishimori:2010au},
 or via the geometrical construction in higher dimensional scenarios\cite{WF_extdim}.

\begin{acknowledgments}
Work was supported  by the Taiwan NSC under Grant No.\
99-2112-M-007-006-MY3.
\end{acknowledgments}

\end{document}